\documentclass[a4paper,UKenglish,cleveref, autoref, thm-restate]{lipics-v2021}
\usepackage[shortend, vlined]{algorithm2e}
\usepackage{pgfplots}
\usepgfplotslibrary{statistics}
\pgfplotsset{compat=1.18}

\hideLIPIcs  

\title{GridOT -- a discrete optimal transport solver on grids}

\author{Johannes Rauch}
{Ulm University, Institute of Optimization and Operations Research, Germany}
{johannes.rauch@uni-ulm.de}
{https://orcid.org/0000-0002-6925-8830}
{}
\author{Leo Zanotti}
{Ulm University, Institute of Optimization and Operations Research, Germany}
{leo.zanotti@uni-ulm.de}
{https://orcid.org/0009-0001-9695-3812}
{}

\authorrunning{J. Rauch, L. Zanotti} 

\Copyright{Johannes Rauch, Leo Zanotti} 

\ccsdesc[]{Mathematics of computing~Mathematical software~Solvers}

\keywords{Discrete optimal transport}

\supplement{
\href{https://github.com/johannesrauch/GridOT/}{\texttt{https://github.com/johannesrauch/GridOT/}}
}

\acknowledgements{We thank Henning Bruhn-Fujimoto for the introduction to the problem and helpful discussions.}

\nolinenumbers 

\usepackage{tikz}
\tikzset{
	dot/.style = {circle, fill, minimum size=#1,
		inner sep=0pt, outer sep=0pt},
	dot/.default = 5pt
}

\usepackage{lmodern}

\DeclareMathOperator{\supp}{supp}
\DeclareMathOperator{\OPT}{OPT}

\begin{document}

\maketitle

\begin{abstract}
We provide an improved implementation of Schmitzer's sparse multi-scale algorithm~\cite{schmitzer2016sparse} for discrete optimal transport on grids.
We report roughly 2--4 times faster runtimes on the DOTmark benchmark~\cite{schrieber2017dotmark}.
The source code is open source and publicly available.
\end{abstract}

\section{Introduction}\label{sec:intro}
Informally, in the \emph{optimal transport} problem we are given two probability distributions over two sets $X$ and $Y$ and a cost function $c: X \times Y \rightarrow \mathbb{R} \cup \{\infty\}$, and we seek a transport function that transforms one probability function into the other as cheaply as possible.
In \emph{discrete} optimal transport, the given probability distributions are discrete; that is, $X$ and $Y$ are discrete and usually finite sets.
Discrete optimal transport \emph{on grids} means that we can identify $X$ and $Y$ with grids in $\mathbb{R}^d$ and the cost function $c$ is nothing else than the squared Euclidean distance.

Discrete optimal transport on grids is an important special case of the general optimal transport problem, because it has numerous applications in image processing and computer vision.
For instance, Werman et al.~\cite{werman1985distance} use a discrete optimal transport problem to define a distance metric for multidimensional histograms.
Building on this, Peleg et al.~\cite{peleg1989unified} present a unified treatment of spatial and gray-level resolution in image digitization.
As image pixels are conventionally arranged in two-dimensional grids, their approaches rely on solving discrete optimal transport on 2-dimensional grids.
Thus, the need for a practically fast solver is evident.

\subsection{Preliminaries}
Before proceeding, we introduce discrete optimal transport on grids formally.
For a finite set $S$ let $\mathcal{P}(S)$ denote the set of probability measures over $S$ with the power set as the $\sigma$-algebra.
Given two finite sets $X$ and $Y$ together with two probability measures $\mu \in \mathcal{P}(X)$ and $\nu \in \mathcal{P}(Y)$, the set of \emph{couplings} or \emph{transport functions} between $\mu$ and $\nu$ is given by
\[
\Pi(\mu,\nu) = \{\pi \in \mathcal{P}(X \times Y): \pi(\{x\} \times Y) = \mu(x), \mu(X \times \{y\}) = \nu(y) \text{ for all $x \in X$, $y \in Y$}\}.
\]
For a cost function $c: X \times Y \rightarrow \mathbb{R} \cup \{\infty\}$ the discrete optimal transport problem is
\begin{align}\tag{$P$}\label{ot:dense}
\min \{C(\pi) : \pi \in \Pi(\mu,\nu)\}, \quad\text{where}\quad C(\pi) = \sum_{(x,y) \in X \times Y} c(x,y)\pi(x,y).
\end{align}
For an integer $i$, let $[i] = \{1, 2, \dots, i\}$.
We say that (\ref{ot:dense}) is on a \emph{grid} if $X = [x_1] \times \dots \times [x_d]$ and $Y = [y_1] \times \dots \times [y_d]$ for some positive integers $d$, $x_1, \dots, x_d$ and $y_1, \dots, y_d$.
Additionally, $c$ is the squared Euclidean distance.
The discrete optimal transport problem (\ref{ot:dense}) is ``dense'' in the sense that an algorithm solving (\ref{ot:dense}) has to consider couplings of all elements in $X \times Y$, of which there are quadratically many.
If $N \subset X \times Y$ is a ``small'' subset, then the restriction
\begin{align}\tag{$P'$}\label{ot:sparse}
\min \{C(\pi) : \pi \in \Pi(\mu,\nu), \supp \pi \subseteq N\}
\end{align}
of (\ref{ot:dense}) is ``sparse'' in the sense that an algorithm solving (\ref{ot:sparse}) only has to consider couplings of all elements in $N$.
We say that $N$ is the \emph{neighborhood} of (\ref{ot:sparse}).
Of course, $\OPT(\text{\ref{ot:dense}}) \leq \OPT(\text{\ref{ot:sparse}})$ holds for any neighborhood $N \subset X \times Y$, and ideally one would want to have a ``small'' neighborhood $N$ such that $\OPT(\text{\ref{ot:dense}}) = \OPT(\text{\ref{ot:sparse}})$.

\subsection{Related work}
Schmitzer devised a sparse multi-scale algorithm for dense optimal transport, which works very well in practice~\cite{schmitzer2016sparse,schrieber2017dotmark}.
Ignoring the multi-scale scheme, we outline his algorithm in Algorithm~\ref{alg:schmitzer}.
The input neighborhood $N$ of Algorithm~\ref{alg:schmitzer} is an ``educated guess'' for a neighborhood $N$ fulfilling $\OPT(\text{\ref{ot:dense}}) = \OPT(\text{\ref{ot:sparse}})$ and usually obtained from the multi-scale scheme.
Algorithm~\ref{alg:schmitzer} repeatedly solves a restricted discrete optimal transport problem (\ref{ot:sparse}) in a neighborhood $N$ and consecutively updates $N$ to solve the discrete optimal transport problem (\ref{ot:dense}) at hand.
For grids, he shows how to construct and update the neighborhoods $N$ in a sparse manner explicitly.
In particular, $\OPT(\text{\ref{ot:dense}}) = \OPT(\text{\ref{ot:sparse}})$ holds for the neighborhood $N$ after the termination of the algorithm.
We refer to his article for a proof of this and more details~\cite{schmitzer2016sparse}.
\begin{algorithm}[h]
\caption{An outline of Schmitzer's algorithm for dense optimal transport~\cite{schmitzer2016sparse} without the multi-scale scheme.}\label{alg:schmitzer}
\DontPrintSemicolon
\KwIn{An instance of discrete optimal transport (\ref{ot:dense}) and a neighborhood $N$.}
\KwOut{An optimal coupling.}
\Repeat{the cost of the coupling does not improve anymore}{
	solve the restricted discrete optimal transport problem (\ref{ot:sparse}) in neighborhood $N$\;
	update $N$\;
}
output the last coupling\;
\end{algorithm}

To solve the restricted discrete optimal transport problem (\ref{ot:sparse}) in Algorithm~\ref{alg:schmitzer}, Schmitzer uses the network simplex algorithm.
An introduction to this algorithm can be found in a book of Ahuja et al.~\cite{ahuja1993networkflows}.
Specifically, he uses the CPLEX\footnote{\href{https://www.ibm.com/products/ilog-cplex-optimization-studio}{\tt https://www.ibm.com/products/ilog-cplex-optimization-studio}} and LEMON~\cite{dezso2010lemon} network simplex implementations for his numerical experiments\footnote{He also uses the cost scaling implementation of LEMON~\cite{dezso2010lemon}, which is outperformed by the network simplex algorithms.}.
Since the problems (\ref{ot:sparse}) are very similar in every iteration of Algorithm~\ref{alg:schmitzer}, it is natural to preserve information in the network simplex algorithm and use warm starts.
While CPLEX provides an interface to set a basis for a warm start, LEMON does not.
To make up for this, Schmitzer uses a trick that modifies the cost function throughout the execution of Algorithm~\ref{alg:schmitzer} with the LEMON network simplex.
Remarkably, he finds that LEMON outperforms CPLEX with this trick~\cite{schmitzer2016sparse}.

\subsection{Our contribution}
We provide a new C++-implementation of Schmitzer's~\cite{schmitzer2016sparse} sparse multi-scale algorithm for dense optimal transport on grids.
Our solver uses a modified network simplex implementation that is adapted from LEMON.
It updates the neighborhood $N$ internally while solving the restricted problem (\ref{ot:sparse}), which preserves information in the network simplex algorithm and eliminates the need for restarts.
Besides, our program distinguishes itself from Schmitzer's in the following points:
We adhere to LEMON's design choice and use compile-time polymorphisms (templates) instead of run-time polymorphisms for efficiency.
Furthermore, we use many of the standard library utilities to avoid dynamically allocating memory and handling raw pointers manually.
This avoids memory leaks and ensures memory safety.
(Using Valgrind\footnote{\href{https://valgrind.org/}{\tt https://valgrind.org/},} we noticed memory leaks in Schmitzer's solver.)
On the DOTmark benchmark~\cite{schrieber2017dotmark}, this results in roughly a 2--4 times faster run-time compared to Schmitzer's solver (see Section~\ref{sec:res}).
Our code is open source and publicly available on GitHub\footnote{\href{https://github.com/johannesrauch/GridOT/}{\texttt{https://github.com/johannesrauch/GridOT/}}}.

\begin{figure}[h]
\caption{The average runtimes in each dataset category and dimension from $32 \times 32$ to $128 \times 128$ together with the speedup factor. The red squares indicate the average runtimes of Schmitzer's solver while the blue dots correspond to the runtimes of our solver. The thick line illustrates the speedup factor.}\label{fig:overview}
\begin{subfigure}{\textwidth}
\centering
\begin{tikzpicture}
\begin{axis}[
xtick=data,
xticklabels={White noise, Rough GRF, Moderate GRF, Smooth GRF, Logarithmic GRF, Logistic GRF, Cauchy density, Shapes, Classic images, Microscopy images},
x tick label style={rotate=75},
xlabel={Dataset category},
axis y line*=left,
ylabel={$t$ in ms},
]
\addplot+ [only marks] table [x=num, y=tms] {data/GridOT-32.txt};
\addplot+ [only marks] table [x=num, y=tms] {data/MultiScaleOT-32.txt};
\end{axis}
\begin{axis}[
hide x axis,
axis y line*=right,
ymin=1,
ymax=4,
ylabel={Speedup factor},
]
\addplot [darkgray, thick] table [x=num, y=speedup] {data/Speedup-32.txt};
\end{axis}
\end{tikzpicture}
\caption{Grid dimension: $32 \times 32$.}
\end{subfigure}
\end{figure}

\begin{figure}[h]
\ContinuedFloat
\begin{subfigure}{\textwidth}
\centering
\begin{tikzpicture}
\begin{axis}[
xtick=data,
xticklabels={White noise, Rough GRF, Moderate GRF, Smooth GRF, Logarithmic GRF, Logistic GRF, Cauchy density, Shapes, Classic images, Microscopy images},
x tick label style={rotate=75},
xlabel={Dataset category},
axis y line*=left,
ylabel={$t$ in ms},
]
\addplot+ [only marks] table [x=num, y=tms] {data/GridOT-64.txt};
\addplot+ [only marks] table [x=num, y=tms] {data/MultiScaleOT-64.txt};
\end{axis}
\begin{axis}[
hide x axis,
axis y line*=right,
ymin=1,
ymax=5,
ylabel={Speedup factor},
]
\addplot [darkgray, thick] table [x=num, y=speedup] {data/Speedup-64.txt};
\end{axis}
\end{tikzpicture}
\caption{Grid dimension: $64 \times 64$.}
\end{subfigure}
\end{figure}

\begin{figure}[h]
\ContinuedFloat
\begin{subfigure}{\textwidth}
\centering
\begin{tikzpicture}
\begin{axis}[
xtick=data,
xticklabels={White noise, Rough GRF, Moderate GRF, Smooth GRF, Logarithmic GRF, Logistic GRF, Cauchy density, Shapes, Classic images, Microscopy images},
x tick label style={rotate=75},
xlabel={Dataset category},
axis y line*=left,
ylabel={$t$ in ms},
]
\addplot+ [only marks] table [x=num, y=tms] {data/GridOT-128.txt};
\addplot+ [only marks] table [x=num, y=tms] {data/MultiScaleOT-128.txt};
\end{axis}
\begin{axis}[
hide x axis,
axis y line*=right,
ymin=1,
ymax=6,
ylabel={Speedup factor},
]
\addplot [darkgray, thick] table [x=num, y=speedup] {data/Speedup-128.txt};
\end{axis}
\end{tikzpicture}
\caption{Grid dimension: $128 \times 128$.}
\end{subfigure}
\end{figure}

\section{Results}\label{sec:res}
We tested both programs on a computer with an Intel(R) Core(TM) i7-8700K CPU @ 3.70GHz and $2 \times 8$ GiB DDR4 DIMM @ 2133 MHz in dual channel mode.
For the test, we use the DOTmark dataset~\cite{schrieber2017dotmark}.
DOTmark consists of several dataset categories: White noise, rough, moderate and smooth Gaussian random field (GRF), logarithmic and logistic GRF, Cauchy density, shapes, classic images, and microscopy images.
Note that these categories can be divided into two parts: One part is randomly generated and the other part is taken from practical applications.
Each category comes in grid dimensions ranging from $32 \times 32$ to $512 \times 512$, increasing by factors of two.
We only go up to dimension $128 \times 128$ in our tests.
We refer the reader to the article of Schrieber et al.~\cite{schrieber2017dotmark} for more details on DOTmark.

We can think of one dataset as a probability measure $\mu$ or $\nu$, respectively, in a discrete optimal transport problem~(\ref{ot:dense}) on grids.
In each dataset category of DOTmark and each grid dimension from $32 \times 32$ to $128 \times 128$, we pair every two distinct datasets, consider them as $\mu$ and $\nu$, and solve the corresponding discrete optimal transport on grids (\ref{ot:dense}) instance with squared Euclidean distance ten times.
For every such pair, we measure the CPU time of every run and get one datapoint by averaging these ten CPU runtimes.
This is shown in Figure~\ref{fig:boxplots} (a)--(j) in Appendix~\ref{app:boxplots}.
Additionally, the average runtime in each dataset category and dimension is shown as an overview in Figure~\ref{fig:overview} (a)--(c).

Unfortunately, Schmitzer's solver~\cite{schmitzer2016sparse} can only handle strictly positive measures $\mu$ and $\nu$, although measures with $\mu(x)=0$ or $\nu(y)=0$ for some $x \in X$ or $y \in Y$, respectively, are common in practical applications as DOTmark shows.
Therefore, we increment each datapoint of each dataset of DOTmark by one to meet this requirement.

\section{Conclusion}
We showed that there is room for improvement in discrete optimal transport solvers, which are still competitive when compared to continuous, numeric solvers~\cite{schrieber2017dotmark}.
We focused on the case where the sets $X$ and $Y$ of (\ref{ot:dense}) are grids and the cost function $c$ is the squared Euclidean distance.
We achieved roughly a 2--4 times faster runtime compared to Schmitzer's program.
It would be nice to have improved solvers for other geometric constellations of $X$ and $Y$ and other cost functions $c$.
For this, we think that the work of Schmitzer~\cite{schmitzer2016sparse} is again a good starting point, since he not only considers grids and squared Euclidean distance, but also $X,Y \subseteq \mathbb{R}^d$ with squared Euclidean distance or strictly convex cost functions, and $X,Y \subseteq S_d = \{x \in \mathbb{R}^d: \Vert x \Vert = 1\}$ with squared geodesic distance.
Another direction for future work is to implement a general interface to support warm starts in LEMON's network simplex algorithm.
As evident from Schmitzer's and this article~\cite{schmitzer2016sparse}, LEMON is a powerful open source library and the network simplex algorithm is an important tool in other algorithms.
Therefore, it would be a good idea to further develop LEMON and unlock its full potential.


\newpage
\appendix
\section{Boxplots}\label{app:boxplots}
\begin{figure}[h]
\caption{Boxplots of the averaged runtimes for each pair in each dataset category and each dimension from $32 \times 32$ to $128 \times 128$. The runtimes of Schmitzer's solver are red while the runtimes of our solver are blue.}\label{fig:boxplots}
\begin{subfigure}{0.4\textwidth}
\begin{tikzpicture}[scale=0.8]
\begin{axis}[
boxplot/draw direction=y, 
boxplot/box extend=0.25,
xtick={5,6,7}, 
ymode=log,
xlabel={$\log_2(\dim)$},
ylabel={$t$ in ms},
]
\input{data/GridOT-WhiteNoise-32.txt}
\input{data/GridOT-WhiteNoise-64.txt}
\input{data/GridOT-WhiteNoise-128.txt}
\input{data/MultiScaleOT-WhiteNoise-32.txt}
\input{data/MultiScaleOT-WhiteNoise-64.txt}
\input{data/MultiScaleOT-WhiteNoise-128.txt}
\end{axis}
\end{tikzpicture}
\caption{White noise.}
\end{subfigure}
\hspace{12mm}
\begin{subfigure}{0.4\textwidth}
\begin{tikzpicture}[scale=0.8]
\begin{axis}[
boxplot/draw direction=y, 
boxplot/box extend=0.25,
xtick={5,6,7}, 
ymode=log,
xlabel={$\log_2(\dim)$},
ylabel={$t$ in ms},
]
\input{data/GridOT-GRFrough-32.txt}
\input{data/GridOT-GRFrough-64.txt}
\input{data/GridOT-GRFrough-128.txt}
\input{data/MultiScaleOT-GRFrough-32.txt}
\input{data/MultiScaleOT-GRFrough-64.txt}
\input{data/MultiScaleOT-GRFrough-128.txt}
\end{axis}
\end{tikzpicture}
\caption{Rough GRF.}
\end{subfigure}
\end{figure}

\begin{figure}[h]
\ContinuedFloat
\begin{subfigure}{0.4\textwidth}
\begin{tikzpicture}[scale=0.8]
\begin{axis}[
boxplot/draw direction=y, 
boxplot/box extend=0.25,
xtick={5,6,7}, 
ymode=log,
xlabel={$\log_2(\dim)$},
ylabel={$t$ in ms},
]
\input{data/GridOT-GRFmoderate-32.txt}
\input{data/GridOT-GRFmoderate-64.txt}
\input{data/GridOT-GRFmoderate-128.txt}
\input{data/MultiScaleOT-GRFmoderate-32.txt}
\input{data/MultiScaleOT-GRFmoderate-64.txt}
\input{data/MultiScaleOT-GRFmoderate-128.txt}
\end{axis}
\end{tikzpicture}
\caption{Moderate GRF.}
\end{subfigure}
\hspace{12mm}
\begin{subfigure}{0.4\textwidth}
\begin{tikzpicture}[scale=0.8]
\begin{axis}[
boxplot/draw direction=y, 
boxplot/box extend=0.25,
xtick={5,6,7}, 
ymode=log,
xlabel={$\log_2(\dim)$},
ylabel={$t$ in ms},
]
\input{data/GridOT-GRFsmooth-32.txt}
\input{data/GridOT-GRFsmooth-64.txt}
\input{data/GridOT-GRFsmooth-128.txt}
\input{data/MultiScaleOT-GRFsmooth-32.txt}
\input{data/MultiScaleOT-GRFsmooth-64.txt}
\input{data/MultiScaleOT-GRFsmooth-128.txt}
\end{axis}
\end{tikzpicture}
\caption{Smooth GRF.}
\end{subfigure}
\end{figure}

\begin{figure}[h]
\ContinuedFloat
\begin{subfigure}{0.4\textwidth}
\begin{tikzpicture}[scale=0.8]
\begin{axis}[
boxplot/draw direction=y, 
boxplot/box extend=0.25,
xtick={5,6,7}, 
ymode=log,
xlabel={$\log_2(\dim)$},
ylabel={$t$ in ms},
]
\input{data/GridOT-LogGRF-32.txt}
\input{data/GridOT-LogGRF-64.txt}
\input{data/GridOT-LogGRF-128.txt}
\input{data/MultiScaleOT-LogGRF-32.txt}
\input{data/MultiScaleOT-LogGRF-64.txt}
\input{data/MultiScaleOT-LogGRF-128.txt}
\end{axis}
\end{tikzpicture}
\caption{Logarithmic GRF.}
\end{subfigure}
\hspace{12mm}
\begin{subfigure}{0.4\textwidth}
\begin{tikzpicture}[scale=0.8]
\begin{axis}[
boxplot/draw direction=y, 
boxplot/box extend=0.25,
xtick={5,6,7}, 
ymode=log,
xlabel={$\log_2(\dim)$},
ylabel={$t$ in ms},
]
\input{data/GridOT-LogitGRF-32.txt}
\input{data/GridOT-LogitGRF-64.txt}
\input{data/GridOT-LogitGRF-128.txt}
\input{data/MultiScaleOT-LogitGRF-32.txt}
\input{data/MultiScaleOT-LogitGRF-64.txt}
\input{data/MultiScaleOT-LogitGRF-128.txt}
\end{axis}
\end{tikzpicture}
\caption{Logistic GRF.}
\end{subfigure}
\end{figure}

\begin{figure}[h]
\ContinuedFloat
\begin{subfigure}{0.4\textwidth}
\begin{tikzpicture}[scale=0.8]
\begin{axis}[
boxplot/draw direction=y, 
boxplot/box extend=0.25,
xtick={5,6,7}, 
ymode=log,
xlabel={$\log_2(\dim)$},
ylabel={$t$ in ms},
]
\input{data/GridOT-CauchyDensity-32.txt}
\input{data/GridOT-CauchyDensity-64.txt}
\input{data/GridOT-CauchyDensity-128.txt}
\input{data/MultiScaleOT-CauchyDensity-32.txt}
\input{data/MultiScaleOT-CauchyDensity-64.txt}
\input{data/MultiScaleOT-CauchyDensity-128.txt}
\end{axis}
\end{tikzpicture}
\caption{Cauchy density.}
\end{subfigure}
\hspace{12mm}
\begin{subfigure}{0.4\textwidth}
\begin{tikzpicture}[scale=0.8]
\begin{axis}[
boxplot/draw direction=y, 
boxplot/box extend=0.25,
xtick={5,6,7}, 
ymode=log,
xlabel={$\log_2(\dim)$},
ylabel={$t$ in ms},
]
\input{data/GridOT-Shapes-32.txt}
\input{data/GridOT-Shapes-64.txt}
\input{data/GridOT-Shapes-128.txt}
\input{data/MultiScaleOT-Shapes-32.txt}
\input{data/MultiScaleOT-Shapes-64.txt}
\input{data/MultiScaleOT-Shapes-128.txt}
\end{axis}
\end{tikzpicture}
\caption{Shapes.}
\end{subfigure}
\end{figure}

\begin{figure}[h]
\ContinuedFloat
\begin{subfigure}{0.4\textwidth}
\begin{tikzpicture}[scale=0.8]
\begin{axis}[
boxplot/draw direction=y, 
boxplot/box extend=0.25,
xtick={5,6,7}, 
ymode=log,
xlabel={$\log_2(\dim)$},
ylabel={$t$ in ms},
]
\input{data/GridOT-ClassicImages-32.txt}
\input{data/GridOT-ClassicImages-64.txt}
\input{data/GridOT-ClassicImages-128.txt}
\input{data/MultiScaleOT-ClassicImages-32.txt}
\input{data/MultiScaleOT-ClassicImages-64.txt}
\input{data/MultiScaleOT-ClassicImages-128.txt}
\end{axis}
\end{tikzpicture}
\caption{Classic images.}
\end{subfigure}
\hspace{12mm}
\begin{subfigure}{0.4\textwidth}
\begin{tikzpicture}[scale=0.8]
\begin{axis}[
boxplot/draw direction=y, 
boxplot/box extend=0.25,
xtick={5,6,7}, 
ymode=log,
xlabel={$\log_2(\dim)$},
ylabel={$t$ in ms},
]
\input{data/GridOT-MicroscopyImages-32.txt}
\input{data/GridOT-MicroscopyImages-64.txt}
\input{data/GridOT-MicroscopyImages-128.txt}
\input{data/MultiScaleOT-MicroscopyImages-32.txt}
\input{data/MultiScaleOT-MicroscopyImages-64.txt}
\input{data/MultiScaleOT-MicroscopyImages-128.txt}
\end{axis}
\end{tikzpicture}
\caption{Microscopy images.}
\end{subfigure}
\end{figure}
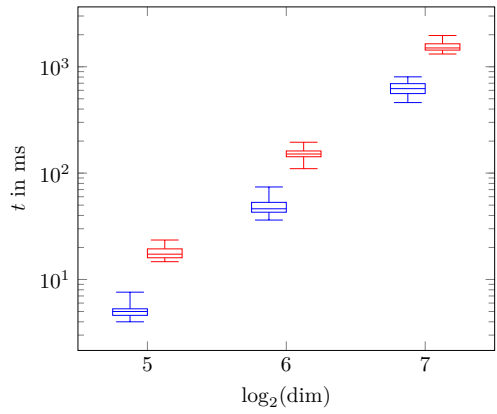
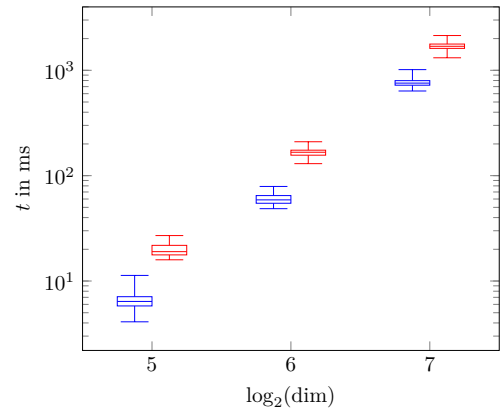
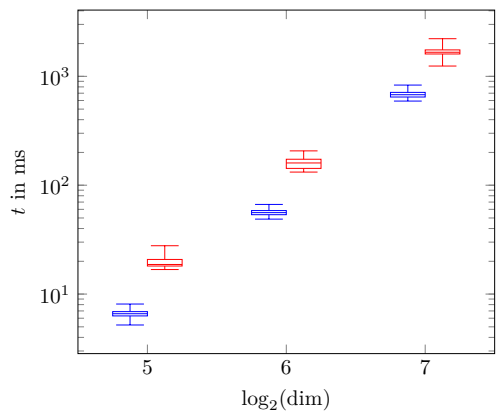
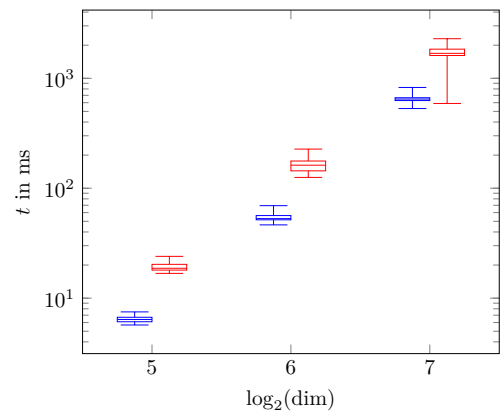
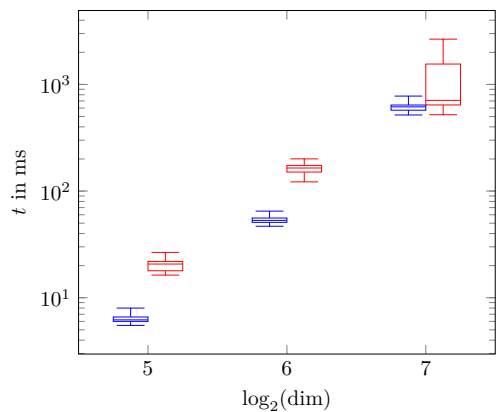
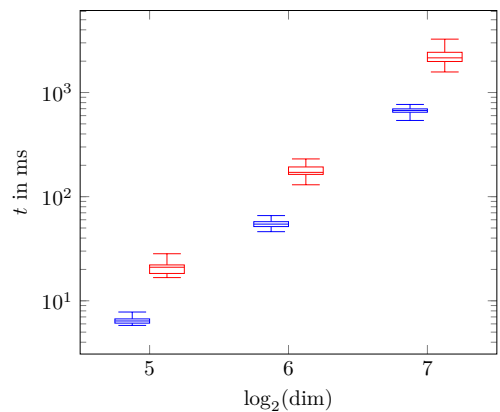
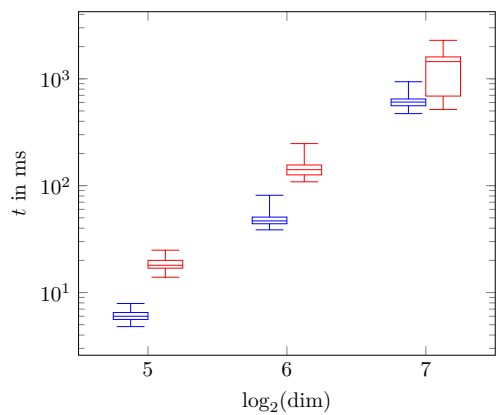
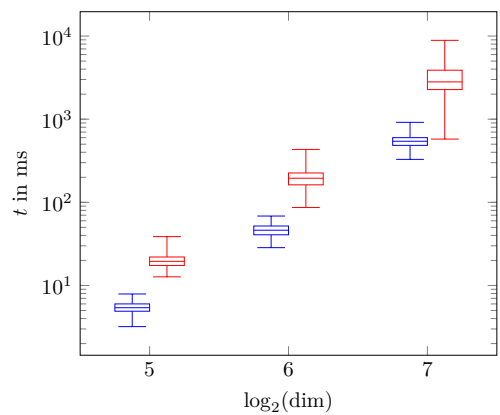
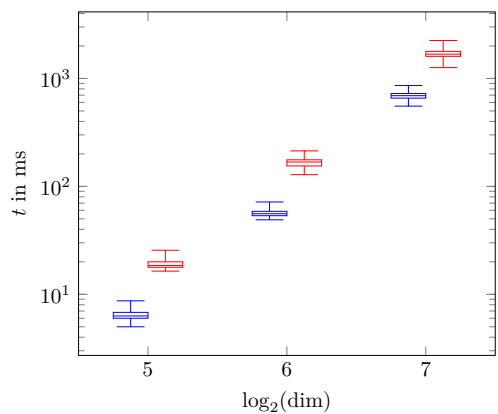
\end{document}